\documentclass[10pt,twocolumn,english,aps]{revtex4}
\usepackage[T1]{fontenc}
\usepackage[latin9]{inputenc}
\usepackage{color}
\usepackage{array}
\usepackage{amsmath}
\usepackage{graphicx}
\usepackage{amssymb}
\usepackage{epstopdf}
\usepackage{float}
\restylefloat{table}
\makeatletter

\providecommand{\tabularnewline}{\\}

\usepackage{epsfig}\usepackage{bm}\usepackage{epsfig}\@ifundefined{definecolor}
 {\usepackage{color}}{}

\def\be{\begin{equation}}
\def\ee{\end{equation}}
\def\bea{\begin{eqnarray}}
\def\eea{\end{eqnarray}}

\@ifundefined{showcaptionsetup}{}{%
 \PassOptionsToPackage{caption=false}{subfig}}
\usepackage{subfig}
\makeatother

\usepackage{babel}

\begin{document}

\title{Direct quantum communication
without actual transmission of the message qubits}

\author{Chitra Shukla$^{1}$  and  Anirban Pathak$^{1,2}$}

\affiliation{$^{1}$Jaypee Institute of Information Technology, A-10, Sector-62,
Noida, UP-201307, India\\
$^{2}$RCPTM, Joint Laboratory of Optics of Palacky University and
Institute of Physics of Academy of Science of the Czech Republic,
Faculty of Science, Palacky University, 17. listopadu 12, 771 46
Olomouc, Czech Republic}

\begin{abstract}
Recently an orthogonal state based protocol of direct quantum communication
without actual transmission of particles is proposed by Salih \emph{et
al. }{[}Phys. Rev. Lett. \textbf{110} (2013) 170502{]} using chained
quantum Zeno effect. As the no-transmission of particle claim is criticized
by Vaidman {[}arXiv:1304.6689 (2013){]}, the condition (claim) of
Salih \emph{et al.} is weaken here to the extent that transmission
of particles is allowed, but transmission of the message qubits (the
qubits on which the secret information is encoded) is not allowed.
Remaining within this weaker condition it is shown that there exists a large
class of quantum states, that can be used to implement an orthogonal
state based protocol of secure direct quantum communication using
entanglement swapping, where actual transmission of the message qubits
is not required. The security of the protocol originates from monogamy of entanglement.
As the protocol can be implemented without using conjugate coding
its security is independent of non-commutativity. 
\end{abstract}


\maketitle

\section{Introduction\label{sec:Introduction}}

The need of secrecy is eternal. From the beginning of human civilization
several methods of secure communication have been proposed and implemented.
However, until the appearance of quantum cryptography, none of the
methods/protocols of secure communication was unconditionally secure.
First ever unconditionally secure protocol of quantum key distribution
(QKD) was introduced by Bennett and Brassard in 1984 \cite{bb84}.
Since this pioneering work on quantum cryptography several other protocols
of secure quantum communication have been proposed to date. Interestingly,
applicability of all the early protocols of secure quantum communication
\cite{bb84,ekert,b92,vaidman-goldenberg}, were limited to QKD. However,
it was realized very soon that quantum states can be employed for
other more general cryptographic tasks, too. For example, in 1999
Hillery \emph{et al.} introduced a protocol for quantum secret sharing
(QSS) of classical secrets \cite{Hillery}. Almost simultaneously,
a protocol for \textit{deterministic secure quantum communication}
(DSQC) was proposed by Shimizu and Imoto \cite{Imoto}. Eventually
the protocol of Shimizu and Imoto was found to be insecure in its
original form.\textcolor{red}{{} }However, it suggested that there
may exists a protocol of quantum communication that can circumvent
prior generation of keys (i.e., QKD) and thus it may lead to unconditionally
secure direct quantum communication. Subsequently, many such protocols
have been proposed. Such protocols can broadly be divided into two
classes: (a) protocols for quantum secure direct communication (QSDC)
\cite{Long   and   Liu,ping-pong,for PP,lm05} and (b) protocols for
DSQC \cite{dsqc_summation,dsqcqithout-max-entanglement,dsqcwithteleporta,entanglement      swapping,Hwang-Hwang-Tsai,reordering1,the:cao and song,the:high-capacity-wstate}.

In a secure direct quantum communication protocol if receiver (Bob)
requires a \textit{pre-key} to read out the secret message sent by
the sender (Alice) then the protocol is referred to as DSQC protocol,
otherwise (i.e., if no such pre-key is required) the protocol is referred
to as QSDC protocol. Specifically, in a DSQC protocol Bob can decode
the secret message sent by Alice only after receipt of additional
classical information of at least one bit for each qubit transmitted
by Alice \cite{review}. In the present paper we aim to design a new
orthogonal state based DSQC protocol. Before we do so it would be
interesting to note that all DSQC protocols can be transformed to
protocols of QKD. It is easy to convert a QSDC/DSQC protocol into
a protocol of QKD by using a random number generator. Specifically,
if we assume that the sender (Alice) possesses a random number generator
and she transmits the outcome of the random number generator to Bob,
then any QSDC/DSQC protocol would reduce to a protocol of QKD. Thus
if we can establish that an orthogonal state based DSQC protocol can
be realized by using any member of a family of quantum states, then
that would imply that QKD can also be realized using these quantum
states. Further, in a conventional QKD protocol we generate an unconditionally
secure key by quantum means and subsequently use classical cryptographic
resources to encode the secret message. No such classical means are
required in DSQC and QSDC. This makes these protocols purely quantum
mechanical. 

The unconditional security of majority of the existing protocols of
DSQC, QSDC and QKD arise from the conjugate coding. Only recently
we have shown that conjugate coding is not essential for DSQC \cite{With preeti,With chitra IJQI}.
Subsequently,  Salih \emph{et al.} have also proposed an orthogonal state based protocol of DSQC  \cite{PRL-counterfactuladsqc}.
Here we provide another orthogonal state based protocol of DSQC. The protocol presented here is a Goldenberg-Vaidman (GV)
type \cite{vaidman-goldenberg} protocol of DSQC, that uses only orthogonal
states for encoding, decoding and error checking, as was done in the
original GV protocol of QKD. Interestingly, GV protocol was introduced
in  1995, but for many years it remained isolated as the only orthogonal-state-based
protocol of QKD. Finally in 2009 another orthogonal state based protocol
known as N09 protocol \cite{N09} was proposed by T.-G. Noh. These
two orthogonal state based protocols are fundamentally different from
the other conjugate coding based (BB84 type) protocols for several
reasons. Most importantly, security of these two protocols does not
depend on noncommutativity.  Consequently, they are extremely important
from the foundational perspectives. Importance of orthogonal-state-based 
protocols are not limited to foundational aspects only, they
are also of practical importance as they are experimentally realizable
\cite{GV-experiment,N09-expt-1,N09 expt-2,N09 expt-3}. To be precise,
recently GV protocol is experimentally realized \cite{GV-experiment}.
A set of successful implementation of N09 protocol are also reported
\cite{N09-expt-1,N09 expt-2,N09 expt-3}. The foundational importance
and the recent experimental achievements have motivated us to investigate
the power of orthogonal state based protocols from various aspects.
Recently, we have shown that the security of both GV and N09 protocols
arise from duality \cite{With   preeti}. We have also generalized
the GV protocol to GV-type DSQC and QSDC protocols. Our Bell-state-based
generalizations of original GV protocol may also be regarded as the
first instance of GV-type DSQC and QSDC protocols \cite{With   preeti}.
We have also shown that maximally efficient orthogonal state based
protocol of DSQC and QSDC can be designed with arbitrary quantum states
\cite{With chitra IJQI}. The foundational importance and the recent
experimental achievements have also motivated others to generalize
N09 protocol to a protocol of secure direct quantum communication.
To be precise, recently Salih \emph{et al.} have provided a very interesting
protocol of direct counterfactual quantum communication \cite{PRL-counterfactuladsqc}
using chained quantum Zeno effect. They claimed that direct quantum
communication between Alice and Bob is possible without actual transmission
of particles between them. This claim is criticized by Vaidman \cite{vaidman-criticism},
who argued that actual measurement of the presence of the qubits in
transmission channel contradicts the claim of Salih \emph{et al.}
The argument of Vaidman motivates us to weaken the claim of Salih
\emph{et al.} (i.e., the condition imposed by Salih \emph{et al.})
and to investigate the possibility of designing a protocol of DSQC
under the condition that transmission of particles are allowed but
transmission of information encoded qubits are not allowed. This condition is referred to as weak condition. Remaining
within this weak condition we have proposed an entanglement swapping based
protocol of DSQC.

In the protocol proposed in this paper, the rearrangement of order
of particles, plays an important role. DSQC protocol based on this
technique was first proposed by Zhu \emph{et al}. \cite{reordering1}
in 2006. However, it was found to be insecure under a Trojan-horse attack,
and was corrected  by Li \emph{et al}. \cite{dsqcqithout-max-entanglement}.
The Li \emph{et al.} protocol may thus be considered as the first
unconditionally secure protocol of DSQC based on \textit{permutation
of particles} (PoP). Since the work of Li \emph{et al.}, many PoP-based
protocols of DSQC have been proposed. Very recently, Banerjee and
Pathak \cite{With   Anindita-pla}, Shukla, Banerjee and Pathak \cite{With chitra-ijtp},
Shukla, Pathak and Srikanth \cite{With chitra IJQI}, Yuan \emph{et
al}. \cite{the:high-capacity-wstate} and Tsai \emph{et al}. \cite{the:C.-W.-Tsai}
have proposed PoP-based protocols for both DSQC and QSDC. The Yuan
\emph{et al}. protocol and Shukla-Banerjee-Pathak protocol use 4-qubit
symmetric $W$ state for communication, while the Banerjee-Pathak
protocol uses 3-qubit $GHZ$-like states, Shukla-Pathak-Srikanth\textcolor{red}{{}
}protocol uses arbitrary quantum states \cite{With chitra IJQI},
and the Tsai \emph{et al.} protocol utilizes the dense coding of four-qubit
cluster states. In all these protocols the qubits on which Alice encodes
a message travel through the quantum channel. In contrast, no such transmission happens in recently 
proposed Zhang \emph{et al.} \cite{QSDC new-cluster} protocol and Salih  \emph{et al.} protocol \cite{PRL-counterfactuladsqc}.
 Extending  their  ideas \cite{PRL-counterfactuladsqc,QSDC new-cluster}
we aim to show that there exists a class of quantum states
that may
be used to implement GV type protocol of DSQC that would be free from
transmission of information encoded qubits. 

The remaining part of the present paper is organized as follows, in
Section \ref{sec:General-form-of} we describe the general form of
a set of quantum states that may be used for DSQC using entanglement
swapping. In this section it is indicated that this set of quantum
states can be used to design entanglement swapping based protocol
of DSQC. A set of well known quantum states are also shown to be member
of this set of quantum states. In Section \ref{sec:Protocol}, we
describe a general GV type orthogonal state based protocol of DSQC
that can be implemented using any quantum state of the family described
in Section \ref{sec:General-form-of}. An explicit example of the
protocol is also provided using $GHZ$-like state. In Section \ref{sec:Security-and-efficiency}
security and efficiency of the protocol is discussed and finally the
paper is concluded in Section \ref{sec:Conclusions}.

\section{General form of the quantum state\label{sec:General-form-of}}

We are interested to design a protocol of DSQC that can transmit an $n$-bit message
using the quantum states of the form \begin{equation}
|\psi\rangle=\frac{1}{\sqrt{2^{n}}}\sum_{i=1}^{2^{n}}|e_{i}\rangle|f_{i}\rangle,\label{eq:state of interest}\end{equation}
where $\left\{ |e_{i}\rangle\right\} $ is a basis set in $C^{2^{m}}:\, m\geq n$
and each of the basis vector is $m$ qubit maximally entangled state
(consequently, $m\geq2)$, and $\left\{ |f_{i}\rangle\right\} $ is
a basis set in $C^{2^{l}}:l\geq n\geq1$. It is not essential for
the basis elements of $\left\{ |f_{i}\rangle\right\} $ to be in entangled
state. Thus $|\psi\rangle$ is a $m+l$ qubit state. Since $\left\{ |e_{i}\rangle\right\} $
and $\left\{ |f_{i}\rangle\right\} $ are basis set, $i\neq i^{\prime}$
implies that $|e_{i}\rangle\neq|e_{i^{\prime}}\rangle$ and $|f_{i}\rangle\neq|f_{i^{\prime}}\rangle$.
This in turn ensures that $|\psi\rangle$ is an entangled state. In
general we demand $|e_{i}\rangle$ as maximally entangled $m$-qubit
state. However, for the convenience of proof we restrict ourselves
to a specific case where $|e_{i}\rangle$ is a $m$-qubit cat state.
Now we assume that the quantum state $|\psi\rangle$ described in
(\ref{eq:state of interest}) is prepared by Alice. She keeps first
$m$ qubits with herself and sends the remaining $l$ qubits to Bob
in a non-clonable manner. By non-clonable manner we mean that Alice
sends the qubits to Bob in such a way that Eve cannot clone the state
$|f_{i}\rangle$. Meaning of this will be elaborated in what follows.
Further, imagine that Alice has prepared another cat state $|e_{j}\rangle$
of same dimension, then the combined state $|e_{j}\rangle\otimes|\psi\rangle$
can be expressed as \begin{equation}
|\psi_{1}\rangle=\frac{1}{\sqrt{2^{n}}}\sum_{i=1}^{2^{n}}|e_{j}\rangle|e_{i}\rangle|f_{i}\rangle=|e_{j}\rangle\left(\frac{1}{\sqrt{2^{n}}}\sum_{i=1}^{2^{n}}|e_{i}\rangle|f_{i}\rangle\right).\label{eq:combined1}\end{equation}

In what follows we will see that Alice encodes a secret $j$ by creating
$|e_{j}\rangle.$ Thus the index $j$ corresponds to a secret bit
string indexed by $j$. From (\ref{eq:combined1}), it is clear that
in $|\psi_{1}\rangle$ the first $m$ qubits (i.e., the qubits of
$|e_{j}\rangle$) are separable from the rest of the qubits. Consequently,
any measurement on rest of the qubits will not reveal any information
about the state of the first $m$ qubits. Let us see what happens
if we allow Alice to do entanglement swapping among first $2m$ qubits
of this combined state (\ref{eq:combined1}). Specifically, we are
interested to see the effect of entanglement swapping on $|e_{j}\rangle|e_{i}\rangle$.
To do so, Alice may follow the following prescription. She takes first
$p=\frac{m}{2}$ qubits from both the cat states (i.e., $|e_{j}\rangle$
and $|e_{i}\rangle$) if $m$ is even, otherwise she takes $p=\frac{m-1}{2}$
qubits from both $|e_{j}\rangle$ and $|e_{i}\rangle$. Thus Alice
has a set of $2p$ qubits (this set is referred to as first set) and
another set of $2\left(m-p\right)$ qubits (this set is referred to
as second set). 

Before we proceed further, the notations used here can be made more
precise using the convention used in \cite{Saugato}. Following \cite{Saugato}
we can express an $m$-qubit cat sate in general as

\begin{equation}
|e\rangle=\frac{1}{\sqrt{2}}\left(\prod_{k=1}^{m}|u_{k}\rangle\pm\prod_{k=1}^{m}|u_{k}^{c}\rangle\right),\label{eq:cat state}\end{equation}
where the symbols $u_{k}$ stands for binary variable $\in\{0,1\}$
with $u_{k}^{c}=1-u_{k}$. The state of our interest (\ref{eq:combined1})
is a finite superposition of products of two cat states. Let each
of these cat sates be labeled by $q$, where $q=1,2$, and the $k$th
particle of the $l$th cat state is labeled by $k(l)$. This summarizes
the notation used here. Now it is straightforward to recognize that
the set of all cat states of $2p$ qubits forms a complete orthonormal
basis set and using the notation described here the elements of such
a basis set can be expressed as \begin{equation}
|\psi(2p)\rangle=\prod_{q=1}^{2}\prod_{k=1}^{p}|u_{k(q)}\rangle\pm\prod_{q=1}^{2}\prod_{k=1}^{p}|u_{k(q)}^{c}\rangle.\label{eq:basis2p}\end{equation}

Now we imagine that Alice performs a projective measurement on the first
set of qubits using cat basis of $2p$ dimension. The measurement
on this basis implies that we operate $|\psi(2p)\rangle\langle\psi(2p)|$
on $|\psi_{1}\rangle.$ The operation would collapse the first set
of qubits into one of the cat states in $2p$ dimension. Remaining
$(2m-2p)$ qubits of $|e_{j}\rangle|e_{i}\rangle$ will be projected
to  a $(2m-2p)$ cat state of the form \cite{Saugato} \begin{equation}
|\psi(2m-2p)\rangle=\prod_{q=1}^{2}\prod_{k=p+1}^{m}|u_{k(q)}\rangle\pm\prod_{q=1}^{2}\prod_{k=p+1}^{m}|u_{k(q)}^{c}\rangle.\label{eq:psi2m-2p}\end{equation}

The structure of (\ref{eq:combined1}) would ensure that the initial
entanglement present between $|e_{i}\rangle$ and $|f_{i}\rangle$
(more precisely between first $m$ particles and last $l$ particles
of $|\psi\rangle)$ is now transferred between the $(2m-2p)$ particles
of $|\psi(2m-2p)\rangle$ and $l$ particles of $|f_{i}\rangle$.


Now if we consider a protocol in which Alice sends the last $l$
qubits (i.e., qubits of $|f_{i}\rangle$) to Bob and measures the
first $2p$ qubits in $2p$-qubit cat basis and the remaining $(2m-2p)$
qubits of her possession in $(2m-2p)$-qubit cat basis and announces
the outcomes, then Bob will be able to infer what was $|e_{j}\rangle$
(equivalently the secret encoded by Alice which is indexed by $j$)
by measuring his qubits in $\left\{ |f_{i}\rangle\right\} $ basis
and using the outcomes of Alice's measurement. Thus it leads to a
protocol of direct quantum communication. 

At a first glance any protocol designed along the above line of arguments
does not appear to be secure as $\left\{ |f_{i}\rangle\right\} $
is orthogonal and measurement outcomes of Alice are public knowledge.
Conventionally, orthogonal states can be perfectly measured and thus
cloned. A measurement by Eve in $\left\{ |f_{i}\rangle\right\} $
basis will destroy the entanglement, but Alice and Bob will not be
able to trace Eve if they apply the above idea without using any strategy
for eavesdropping check. Further, if Eve is allowed to measure
the states communicated by Alice in $\left\{ |f_{i}\rangle\right\} $
basis then she is also capable to clone the states \cite{TMor} and
the protocol would fail. However, it is possible to design strategy
in which orthogonal states are communicated in such a way that Eve
does not have access to the basis set in which the communicated states
are basis elements (i.e., the basis set in which the communicated
states are perfectly measurable). This restriction on the basis sets
available to Eve implies nocloning \cite{TMor} and when orthogonal
states are communicated using such a strategy then we say that the
states are communicated in a non-clonable manner. To communicate the
orthogonal states of $\left\{ |f_{i}\rangle\right\} $ basis in a
non-clonable manner we need to ensure that Eve does not have access
to $\left\{ |f_{i}\rangle\right\} $ basis. This is possible in several
ways. For example, non-clonable communication is possible if the physical realization of all the states in
$\left\{ |f_{i}\rangle\right\} $ basis are such that they may be
visualized as superposition of two or more pieces that can be geographically separated.
For example, in the original GV protocol \cite{vaidman-goldenberg},
orthogonal states $|\phi_{0}\rangle=|a\rangle+|b\rangle$ and $|\phi_{1}\rangle=|a\rangle-|b\rangle$
are used to communicate bit values $0$ and $1$, respectively, but
Alice used to send the wave packet $|b\rangle$ to Bob only after
wavepacket $|a\rangle$ is received by Bob. This strategy implies
that Eve does not have simultaneous access to $|a\rangle$ and $|b\rangle$
and as a consequence Eve cannot perform a measurement in $\{|a\rangle+|b\rangle,\,|a\rangle-|b\rangle\}$
basis. Eve's inability to perform a measurement in $\{|a\rangle+|b\rangle,\,|a\rangle-|b\rangle\}$
basis implies that she can neither do a perfect measurement nor perform
cloning operation \cite{TMor}. Thus in the GV protocol orthogonal
states are communicated in a non-clonable manner.   We are not interested
to follow original GV idea to communicate $|f_{i}\rangle$ in a non-clonable
manner as GV idea requires strict time checking which is difficult
to achieve experimentally. Some of the present authors \cite{With chitra IJQI, With preeti} have
recently generalized the GV idea and have suggested another strategy
of non-clonable communication of orthogonal states by using the fact
that entangled states are nothing but superposition in tensor product
space. In our procedure strict time checking is not required \cite{With preeti}.
To be precise, Alice can concatenate a set of decoy qubits prepared in
 Bell states (say Alice prepares $|\psi^{+}\rangle^{\otimes \frac{N}{2}}=\left(\frac{|00\rangle+|11\rangle}{\sqrt{2}}\right)^{\otimes \frac{N}{2}}$)
 with a $N$-qubit string that she wants to transmit to Bob
and randomly rearrange the particle ordering i.e., apply PoP
technique and thus restrict the basis available to Eve. PoP will ensure
that Eve cannot clone or measure the decoy qubits as she does not
know which qubits are mutually entangled. Further Eve will not be
able to selectively clone or measure non-decoy qubits as after application
of PoP, she has no way to isolate decoy qubits from the other qubits. As
perfect measurement by Eve is not possible due to unavailability of
$\left\{ |f_{i}\rangle\right\} $ basis, any measurement and/or cloning
attempt by Eve will leave a signature, that can be traced by measuring
and comparing the decoy qubits. In summary, Alice can always communicate
last $l$ qubits of (\ref{eq:state of interest}) to Bob in a non-clonable
manner and that in turn  ensures protection against measurement
and resend attack and cloning (CNOT) attack. Above facts lead us to
a protocol of DSQC using entanglement swapping where actual transmission
of the information encoded particles are not required. The protocol
is elaborated on Section \ref{sec:Protocol}.

\subsection{Some special cases of the quantum state}

Till now it may not be very easy to visualize: How general the state
(\ref{eq:state of interest}) is? So here we note some special cases
of this general state (\ref{eq:state of interest}). Let us start
with the simplest case $m=2,\, n=1$ and $l=1$. For $m=2,$ the obvious
choice of maximally entangled basis set is Bell basis $\{|e_{i}\rangle\}=\{|\psi^{+}\rangle,|\psi^{-}\rangle,|\phi^{+}\rangle,|\phi^{-}\rangle\}$, where
$\psi^{\pm}=\frac{|00\rangle \pm|11\rangle}{\sqrt{2}}$ and $\phi^{\pm}=\frac{|01\rangle \pm|10\rangle}{\sqrt{2}}$.
For $n=1,$ we need only 2 basis vectors from the set $\{|e_{i}\rangle\}$.
Let us choose $|e_{1}\rangle=|\psi^{+}\rangle,|e_{2}\rangle=|\psi^{-}\rangle$,
now if we choose $|f_{1}\rangle=|0\rangle$ and $|f_{2}\rangle=|1\rangle,$
then $|\psi\rangle=\frac{1}{\sqrt{2}}\left(|\psi^{+}0\rangle+|\psi^{-}1\rangle\right)$,
which is a $GHZ$-like state \cite{With Anindita IJQI 2011}. Here we
can easily recognize that all $GHZ$-like states are actually of the
form (\ref{eq:state of interest}). Alternatively, if we choose $|e_{1}\rangle=|\psi^{+}\rangle,\,|e_{2}\rangle=|\psi^{-}\rangle$
and $|f_{1}\rangle=|+\rangle$ and $|f_{2}\rangle=|-\rangle,$ then
$|\psi\rangle=\frac{1}{\sqrt{2}}\left(|\psi^{+}+\rangle+|\psi^{-}-\rangle\right)=\frac{1}{\sqrt{2}}\left(|000\rangle+|111\rangle\right)$,
which is a $GHZ$ state. Similarly we can show that all other $GHZ$
states are also of the form (\ref{eq:state of interest}). In the
Table \ref{tab:Intersting-quantum-states} we have provided some more
examples of well known quantum states which are of the form (\ref{eq:state of interest}).
Thus if we can show that (\ref{eq:state of interest}) can be used
for DSQC of $n-$bit classical information then that would mean that
we have a large class of states that can be used for DSQC without
actual transmission of message encoded states.

\section{Orthogonal state based protocol of DSQC\label{sec:Protocol} }

The protocol in general works as follows:
\begin{description}
\item [{Step~1:~}] Alice prepares $|\psi\rangle^{\otimes N}.$ She keeps
the first $m$ qubits of each $|\psi\rangle$ with herself and prepares
a sequence $P_{B}$ with the remaining $l$ qubits. Thus $P_{B}$
is a sequence of $Nl$ qubits. 
\item [{Step~2:}] Alice communicates $P_{B}$ to Bob in a non-clonable
manner. To communicate $P_{B}$ in nonclonable manner Alice prepares
$|\psi^{+}\rangle^{\otimes\frac{Nl}{2}}$ as decoy (checking) qubits
and concatenates the qubits into $P_{B}$ to obtain a longer sequence
$P_{B}^{\prime}$, which has total $2Nl$ qubits. Subsequently, Alice
applies a random permutation operator $\Pi_{2Nl}$ on $P_{B}^{\prime}$
to obtain a new sequence $P_{B}^{\prime\prime}=\Pi_{2Nl}P_{B}^{\prime}$
and sends that to Bob.
\item [{Step~3:}] Alice discloses $\Pi_{2Nl}$ (which includes the coordinates
of the Bell pairs) after receiving authentic acknowledgment of receipt
of all the photons from Bob.
\item [{Step~4:}] Bob rearranges the sequence and measures the transmitted
Bell pairs (that are prepared as decoy qubits) in the Bell basis to
determine if they are in the state $|\psi^{+}\rangle$. If the error
detected by Bob is within a tolerable limit, they continue to the
next step. Otherwise, they discard the protocol and restart from \textbf{Step
1}.
\item [{Step~5:}] Alice encodes her $n$-bit message as follows: She encodes
$0_{1}0_{2}\cdots0_{n},\,0_{1}0_{2}\cdots1_{n},\cdots,\,1_{1}1_{2}\cdots1_{n}$
as $|e_{1}\rangle,|e_{2}\rangle,\cdots,|e_{n}\rangle$ respectively
and combines the encoded state with $|\psi\rangle$. Now if Alice
encodes a secret message $j$ then the complete state of the system
is \[
|\psi_{1}\rangle=\frac{1}{\sqrt{2^{n}}}\sum_{i=1}^{2^{n}}|e_{j}\rangle|e_{i}\rangle|f_{i}\rangle,\]
whose first $2m$ qubits are with Alice and the last $l$ qubits are
with Bob. 
\item [{Step~6:}] Alice performs entanglement swapping operation as described
above (see Section \ref{sec:General-form-of}) and announces her measurement outcomes.
\item [{Step~7:}] Bob measures his qubits in $\left\{ |f_{i}\rangle\right\} $
basis. From his measurement outcomes and from the announcement of
Alice, he can decode the information encoded by Alice. 
\end{description}

\subsection{A special case: Implementation of the protocol using $GHZ$-like state\label{sub:A-special-case-ghzlike}}

Assume that Alice and Bob have agreed on the following encoding. If
Alice has to send $0\,(1)$ then she will encode it as $|\psi^{+}\rangle\,\left(|\psi^{-}\rangle\right)$. 
\begin{enumerate}
\item Alice prepares $N$ copies of a $GHZ$-like state $\frac{1}{\sqrt{2}}\left(|\psi^{+}0\rangle+|\psi^{-}1\rangle\right)$,
i.e., Alice prepares $\frac{1}{\sqrt{2^{n}}}\left(|\psi^{+}0\rangle+|\psi^{-}1\rangle\right)^{\otimes N}.$
She keeps the first two photons with herself and prepares a sequence
$P_{B}$ with all the third qubits. 
\item Then Alice prepares $|\psi^{+}\rangle^{\otimes\frac{N}{2}}$ and concatenates
the qubits into $P_{B}$ to obtain a larger sequence $P_{B}^{\prime}$,
which has total $2N$ qubits. Now Alice applies a random permutation
operator $\Pi_{2N}$ on $P_{B}^{\prime}$ to obtain a new sequence
$P_{B}^{\prime\prime}$ and sends that to Bob.
\item Alice discloses $\Pi_{2N}$ (which includes the coordinates of the
Bell pairs) after receiving authentic acknowledgment of receipt of
all the photons from Bob.
\item Bob rearranges the sequence and measures the transmitted Bell pairs
in the Bell basis to determine if they are in the state $|\psi^{+}\rangle$.
If the error detected is within the tolerable limit, they continue
to the next step. Otherwise, they discard the protocol and restart
from \textbf{Step 1}.
\item After confirmation that no eavesdropping has happened, Alice encodes
her message qubit. Now the complete state of the system is $\frac{1}{\sqrt{2}}\left(|\psi^{\pm}\rangle|\psi^{+}0\rangle+|\psi^{\pm}\rangle|\psi^{-}1\rangle\right)_{12345}.$
Here the qubits 1-4 are with Alice and the last qubit is with Bob.
\item Now Alice does Bell measurements on qubits 1,3 and 2,4 and announces
her result.
\item Bob measures his qubit in computational basis. Using his measurement outcome
and the announcement of Alice, he can decode the information
encoded by Alice. 
\end{enumerate}
Let's see how Bob can decode the information. First we assume that
Alice has encoded $0$, then the combined state can be decomposed
as 
\begin{widetext}
\begin{equation}
\begin{array}{ccc}
\frac{1}{\sqrt{2}}\left(|\psi^{+}\psi^{+}0\rangle+|\psi^{+}\psi^{-}1\rangle\right)_{12345} & = & \frac{1}{2\sqrt{2}}\left[\left\{ |\psi^{+}\psi^{+}\rangle+|\phi^{+}\phi^{+}\rangle+|\phi^{-}\phi^{-}\rangle+|\psi^{-}\psi^{-}\rangle\right\} _{1324}|0\rangle_{5}\right.\\
 & + & \left.\left\{ |\psi^{+}\psi^{-}\rangle-|\phi^{+}\phi^{-}\rangle-|\phi^{-}\phi^{+}\rangle+|\psi^{-}\psi^{+}\rangle\right\} _{1324}|1\rangle_{5}\right].\end{array}\label{eq:GHZ-like-0}\end{equation}
Similarly, if Alice encodes $1$, then the combined state is \begin{equation}
\begin{array}{ccc}
\frac{1}{\sqrt{2}}\left(|\psi^{-}\psi^{+}0\rangle+|\psi^{-}\psi^{-}1\rangle\right)_{12345} & = & \frac{1}{2\sqrt{2}}\left[\left\{ |\psi^{+}\psi^{-}\rangle+|\phi^{+}\phi^{-}\rangle+|\phi^{-}\phi^{+}\rangle+|\psi^{-}\psi^{+}\rangle\right\} _{1324}|0\rangle_{5}\right.\\
 & + & \left.\left\{ |\psi^{+}\psi^{+}\rangle-|\phi^{+}\phi^{+}\rangle-|\phi^{-}\phi^{-}\rangle+|\psi^{-}\psi^{-}\rangle\right\} _{1324}|1\rangle_{5}\right]\end{array}\label{eq:GHZ-like-1}\end{equation}

\end{widetext}
Now from the above two equations it is clear that using the announcement
of measurement outcomes of Alice and the outcome of his own measurement
Bob will be able to decode the encoded information. For clarity we
have shown the relation among Alice's measurement outcomes, Bob's
measurement outcomes and Bob's conclusions in Table \ref{tab:Relation-between-Alice's-tab-1}.
Similar expansion and subsequently tables relating Alice's outcome,
Bob's outcome and encoded bit string can be obtained for other quantum
states of the generic form (\ref{eq:state of interest}). For example,
we can easily obtain such tables for all other quantum states listed
in Table \ref{tab:Intersting-quantum-states}.
\begin{widetext}
\begin{table}[H]
\begin{centering}
\begin{tabular}{|>{\centering}p{0.5in}|c|>{\centering}p{1.8in}|>{\centering}p{1.5in}|>{\centering}p{1.8in}|>{\centering}p{0.7in}|}
\hline 
{\small Example No.} & {\small $(l,m,n)$} & {\small $\left\{ |e_{i}\rangle\right\} $} & {\small $\left\{ |f_{i}\rangle\right\} $} & {\small $|\psi\rangle=\frac{1}{\sqrt{2^{n}}}\sum_{i=1}^{2^{n}}|e_{i}\rangle|f_{i}\rangle$} & {\small The state is known as}\tabularnewline
\hline 
{\small 1.} & {\small $(2,2,1)$} & {\small $\{|\psi^{+}\rangle,|\psi^{-}\rangle,|\phi^{+}\rangle,|\phi^{-}\rangle\}$} & {\small $\left\{ |00\rangle,|11\rangle,|01\rangle,|10\rangle\right\} $} & {\small \[
\begin{array}{c}
\frac{1}{\sqrt{2}}\left(|\psi^{+}00\rangle+|\psi^{-}11\rangle\right)\\
=\frac{1}{2}\left(|0000\rangle+|0011\rangle\right.\\
+\left.|1100\rangle-|1111\rangle\right)_{1234}\end{array}\]
} & {\small Cluster state}\tabularnewline
\hline 
{\small 2.} & {\small $(2,2,2)$} & {\small $\{|\psi^{+}\rangle,|\psi^{-}\rangle,|\phi^{+}\rangle,|\phi^{-}\rangle\}$} & {\small $\{|\psi^{-}\rangle,|\psi^{+}\rangle,|\phi^{+}\rangle,|\phi^{-}\rangle\}$} & {\small \[
\begin{array}{c}
\frac{1}{2}\left(|\psi^{+}\psi^{-}\rangle+|\psi^{-}\psi^{+}\rangle\right.\\
+\left.|\phi^{+}\phi^{+}\rangle+|\phi^{-}\phi^{-}\rangle\right)\\
=\frac{1}{2}\left(|0000\rangle+|0101\rangle\right.\\
+\left.|1010\rangle-|1111\rangle\right)_{1324}\end{array}\]
} & {\small Cluster state after swapping of particles 2 and 3}\tabularnewline
\hline 
{\small 3.} & {\small $(2,2,1)$} & {\small $\{|\psi^{+}\rangle,|\psi^{-}\rangle,|\phi^{+}\rangle,|\phi^{-}\rangle\}$} & {\small $\{|\psi^{+}\rangle,|\psi^{-}\rangle,|\phi^{+}\rangle,|\phi^{-}\rangle\}$} & {\small $\frac{1}{\sqrt{2}}\left(|\psi^{+}\psi^{+}\rangle+|\psi^{-}\psi^{-}\rangle\right)$} & {\small 4-qubit cat state}\tabularnewline
\hline 
{\small 4.} & {\small $(1,2,1)$} & {\small $\{|\psi^{+}\rangle,|\psi^{-}\rangle,|\phi^{+}\rangle,|\phi^{-}\rangle\}$} & {\small $\{|+\rangle,|-\rangle\}$} & {\small $\frac{1}{\sqrt{2}}\left(|\psi^{+}+\rangle+|\psi^{-}-\rangle\right)$} & {\small $GHZ$ state}\tabularnewline
\hline 
{\small 5.} & {\small $(1,2,1)$} & {\small $\{|\psi^{+}\rangle,|\psi^{-}\rangle,|\phi^{+}\rangle,|\phi^{-}\rangle\}$} & {\small $\{|0\rangle,|1\rangle\}$} & {\small $\frac{1}{\sqrt{2}}\left(|\psi^{+}0\rangle+|\psi^{-}1\rangle\right)$} & {\small $GHZ$-like state}\tabularnewline
\hline 
{\small 6.} & {\small $(2,3,2)$} & {\small $\begin{array}{c}
\left\{ |G_{010}\rangle,|G_{111}\rangle,|G_{001}\rangle,|G_{100}\rangle,\right.\\
\left.|G_{000}\rangle,|G_{011}\rangle,|G_{101}\rangle,|G_{110}\rangle\right\} \end{array}$} & {\small $\left\{ |00\rangle,-|01\rangle,|10\rangle,-|11\rangle\right\} $} & {\small $\begin{array}{c}
\frac{1}{2}\left(|G_{010}\rangle|00\rangle-|G_{111}\rangle|01\rangle\right.\\
\left.+|G_{001}\rangle|10\rangle-|G_{100}\rangle|11\rangle\right)\\
=|\psi_{B}\rangle_{12534}\end{array}$} & {\small Brown state after swapping of particles}\tabularnewline
\hline 
{\small 7.} & {\small $(2,2,2)$} & {\small $\begin{array}{c}
\left\{ |\Phi_{1}^{+}\rangle=\frac{1}{\sqrt{2}}\left(|\psi^{+}\rangle+|\phi^{-}\rangle\right),\right.\\
\left.|\Phi_{1}^{-}\rangle=\frac{1}{\sqrt{2}}\left(|\psi^{+}\rangle-|\phi^{-}\rangle\right),\right.\\
|\Psi_{1}^{+}\rangle=\frac{1}{\sqrt{2}}\left(|\phi^{+}\rangle+|\psi^{-}\rangle\right),\\
\left.|\Psi_{1}^{-}\rangle=\frac{1}{\sqrt{2}}\left(|\phi^{+}\rangle-|\psi^{-}\rangle\right)\right\} \end{array}$} & {\small $\left\{ |0-\rangle,|0+\rangle,|1-\rangle,|1+\rangle\right\} $} & {\small $\begin{array}{c}
\frac{1}{2}\left(|\Phi_{1}^{+}\rangle|0-\rangle+|\Phi_{1}^{-}\rangle|0+\rangle\right.\\
+\left.|\Psi_{1}^{+}\rangle|1-\rangle+|\Psi_{1}^{-}\rangle|1+\rangle\right)\\
=\frac{1}{2\sqrt{2}}\left(|0000\rangle-|0011\rangle\right.\\
-|0101\rangle+|0110\rangle\\
+|1001\rangle+|1010\rangle\\
\left.+|1100\rangle+|1111\rangle\right)\\
=|\chi\rangle_{1234}\end{array}$} & {\small $\chi$ state}\tabularnewline
\hline 
{\small 8.} & {\small $(2,2,2)$} & {\small $\{|\psi^{+}\rangle,|\psi^{-}\rangle,|\phi^{+}\rangle,|\phi^{-}\rangle\}$} & {\small $\{|\psi^{-}\rangle,|\psi^{+}\rangle,|\phi^{+}\rangle,-|\phi^{-}\rangle\}$} & {\small \[
\begin{array}{c}
\frac{1}{2}\left(|\psi^{+}\psi^{-}\rangle+|\psi^{-}\psi^{+}\rangle\right.\\
+\left.|\phi^{+}\phi^{+}\rangle-|\phi^{-}\phi^{-}\rangle\right)\\
=\frac{1}{2}(|0000\rangle+|0110\rangle\\
+|1001\rangle-|1111\rangle)_{1234}\end{array}\]
} & {\small $\Omega$ state }\tabularnewline
\hline
\end{tabular}
\par\end{centering}

\caption{\label{tab:Intersting-quantum-states}Interesting quantum states of
the form (\ref{eq:state of interest}). Here $|G_{ijk}\rangle=|0\rangle|j\rangle|k\rangle+(-1)^{i}|1\rangle|j\oplus1\rangle|k\oplus1\rangle$,
$i,j,k\in\{0,1\}$ is a $GHZ$ state.}

\end{table}
\end{widetext}

\begin{table}
\begin{centering}
\begin{tabular}{|c|c|c|}
\hline 
Alice's outcome & Bob's outcome & Encoded bit\tabularnewline
\hline 
$\psi^{+}\psi^{+}$ & 0 & 0\tabularnewline
\hline 
$\psi^{+}\psi^{+}$ & 1 & 1\tabularnewline
\hline 
$\psi^{-}\psi^{-}$ & 0 & 0\tabularnewline
\hline 
$\psi^{-}\psi^{-}$ & 1 & 1\tabularnewline
\hline 
$\phi^{+}\phi^{+}$ & 0 & 0\tabularnewline
\hline 
$\phi^{+}\phi^{+}$ & 1 & 1\tabularnewline
\hline 
$\phi^{-}\phi^{-}$ & 0 & 0\tabularnewline
\hline 
$\phi^{-}\phi^{-}$ & 1 & 1\tabularnewline
\hline 
$\psi^{+}\psi^{-}$ & 0 & 1\tabularnewline
\hline 
$\psi^{+}\psi^{-}$ & 1 & 0\tabularnewline
\hline 
$\psi^{-}\psi^{+}$ & 0 & 1\tabularnewline
\hline 
$\psi^{-}\psi^{+}$ & 1 & 0\tabularnewline
\hline 
$\phi^{-}\phi^{+}$ & 0 & 1\tabularnewline
\hline 
$\phi^{-}\phi^{+}$ & 1 & 0\tabularnewline
\hline 
$\phi^{+}\phi^{-}$ & 0 & 1\tabularnewline
\hline 
$\phi^{+}\phi^{-}$ & 1 & 0\tabularnewline
\hline
\end{tabular}
\par\end{centering}

\caption{\label{tab:Relation-between-Alice's-tab-1}Relation between Alice's
outcomes, Bob's outcome and the encoded information.}
\end{table}

\section{Security and efficiency of the protocol\label{sec:Security-and-efficiency}}

In this protocol the qubits with encoded message are not transmitted,
so we just need to ensure that the first transmission (i.e., the transmission
of $l$ qubits of $|f_{i}\rangle$ to Bob) is secure. As described
above, while sending this set of qubits Alice randomly inserts the
decoy qubits prepared in $|\psi^{+}\rangle^{\otimes \frac{Nl}{2}}$ and subsequently
on Alice's disclosure of position of those qubits Bob can measure
them in Bell basis and check what \% of them are still in $|\psi^{+}\rangle$
and use that to compute the error rate. Now we check the protection
of the protocol against some well known eavesdropping strategies. 

\textbf{Measurement and resend attack:} If Eve plans to measure the
qubits of $P_{B}^{\prime\prime}$ and send the outcome to Bob, she will
always leave a signature as she does not know which qubits are entangled
with which one. As Eve has no way to distinguish between decoy qubits and the
other qubits she will end up measuring decoy qubits, clearly
any attempt to measure the decoy qubits in a basis other than Bell
basis will destroy/modify the entanglement and will be traced when
Bob measures them. Even if Eve measures in Bell basis, she will be
traced. This can be visualized through a simple example. Consider
that Eve measures decoy qubits $|\psi^{+}\rangle_{ab}|\psi^{+}\rangle_{cd}$
in Bell basis but as she is unaware of the fact that $a\,(c)$ is
entangled with \textbf{$b\,(d)$}, she measures $ac$ and $bd$ in
Bell basis. Now as
\[
\begin{array}{lcl}
|\psi^{+}\rangle_{ab}|\psi^{+}\rangle_{cd} & = & \frac{1}{2}\left\{ |\psi^{+}\psi^{+}\rangle+|\phi^{+}\phi^{+}\rangle\right.\\
 & + & \left.|\phi^{-}\phi^{-}\rangle+|\psi^{-}\psi^{-}\rangle\right\} _{abcd},\end{array}\] 
75\% of the times measurement of Eve will be detected by Bob as he
will not obtain $|\psi^{+}\rangle_{ab}|\psi^{+}\rangle_{cd}$ as outcome
of his measurement. Similarly other situations can be investigated
to show that the protocol is protected against measurement
and resend attack. 

\textbf{CNOT (Cloning) attack}: PoP ensures that Eve does not have
access to Bell basis. So she cannot try to clone in Bell basis. However
she may try to apply CNOT gate and use each transmitted qubit as control
qubit and prepare target qubits in $|0\rangle$. In such a case her
operation will lead to a decoy state $|\psi^{+}\rangle=\frac{|00\rangle+|11\rangle}{\sqrt{2}}$
to $\frac{|0000\rangle+|1111\rangle}{\sqrt{2}}=\frac{\left(|\psi^{+}\rangle|\psi^{+}\rangle+|\psi^{-}\rangle|\psi^{-}\rangle\right)}{\sqrt{2}}$
where last two qubits are the auxiliary qubits introduced by Eve.
Now 50\% of the time Bob's measurement will yield $|\psi^{-}\rangle$
and Eve will be detected. It is important to note that this is applicable
in general independent of whether elements of $\{|f_{i}\rangle\}$
are entangled or separable as Eve has no way to distinguish between
a decoy qubit and other qubits so she cannot selectively clone.

\textbf{Capture and replacement attack:} In principle Eve can capture
all the qubits sent by Alice and prepare fake states $|\psi\rangle=\frac{1}{\sqrt{2^{n}}}\sum_{i=1}^{2^{n}}|e_{i}\rangle|f_{i}\rangle$
and $|\psi^{+}\rangle^{\otimes \frac{Nl}{2}}$ and use them to prepare a fake
$P_{B}^{\prime\prime}.$ This strategy would fail as the permutation operator
randomly rearranges the qubits, so while Bob will recreate the sequence
after Alices's disclosure of $\Pi_{Nl}$ he will obtain a sequence that
is not same as what was prepared by Eve after concatenating $|\psi^{+}\rangle^{\otimes \frac{Nl}{2}}$
to $P_{B}.$ As a consequence Bob's measuremet outcomes will not always
be $|\psi^{+}\rangle.$ 

We have seen that the protocol is protected against several eavesdropping
attacks. This is expected as the strategy adopted here restricts the
basis set available to Eve. Further, we would like to note that announcements
made by Alice do not disclose any information. This can be visualized
quickly if we note that $|f_{i}\rangle$ can be found in $2^{n}$
different states and as Eve is completely unaware of $|f_{i}\rangle,$
her ignorance is of $\log_{2}2^{n}=n$ bits which is the same as the amount
of information encoded through $|e_{j}\rangle.$ Thus Alice's disclosure
does not reduce the uncertainty of Eve. In the existing entanglement
swapping based DSQC protocols \cite{Ba An} conjugate coding based
techniques that rely on BB84 kind of eavesdropping checking are used.
Thus the security of those protocols essentially arise from noncommutativity.
In contrast to the existing protocols our protocol is completely orthogonal
state based (GV type) protocol and its security arises from monogamy
\cite{With chitra IJQI,With preeti}. 

Qubit efficiency $\eta$ is used for analysis of efficiency of DSQC
and QSDC protocols. It is defined as \cite{defn of qubit efficiency}
\begin{equation}
\eta=\frac{c}{q+b},\label{eq:efficiency 2}\end{equation}
where $c$ denotes the total number of transmitted classical bits
(message bits), $q$ denotes the total number of qubits used and $b$
is the number of classical bits exchanged for decoding of the message
(classical communications used for checking of eavesdropping is not
counted). In our case $|\psi\rangle$ is a $m+l$ qubit state and
in addition we need to add $l$ decoy qubits \cite{With Anindita-pla}
and $m$ qubits for entanglement swapping. So $q=2m+2l$. Now this
state can be used to communicate $n$ bit of classical information
which implies $c=n$. In addition Alice has to disclose her measurement
outcomes by transmitting $2m$ bit classical information. Therefore
\[
\eta=\frac{n}{2m+2l+2m}=\frac{1}{2}\left(\frac{n}{2m+l}\right).\]
Now in the limiting case when $m=n=l$ then the qubit efficiency $\eta=\frac{1}{6}=16.6\%$
is maximum but this value is lower compared to the maximum possible
value of $\eta=\frac{1}{3}=33.33\%$ \cite{With Anindita-pla}. The
difference arises from the definition of $q$, if instead of total
number of qubits we use $q=$ total number of qubits transmitted then
in the limiting case $m=n=l$ the efficiency of the proposed protocol
is 25\%. Still its not maximally efficient. Thus the most interesting feature
of the present protocol that the message encoded states are not transmitted
is associated with a cost as it reduces the efficiency.

\section{Conclusions\label{sec:Conclusions}}

It is shown that DSQC is possible without actual transmission of message
string and the task can be performed with any member of a set of quantum
states having generic form (\ref{eq:state of interest}). The proposed
protocol is based on Bose et al.'s idea of generalized entanglement
swapping \cite{Saugato}, and it is a GV-type orthogonal state based
protocol of DSQC. We have also elaborated the working of the protocol
by considering a special case where the initial state is a $GHZ$-like
state. The protocol is different from most of the conventional DSQC
protocols for the following three reasons: (1) It is an orthogonal
state based protocol and except our recent proposals \cite{With preeti,With chitra IJQI} and Salih \emph{et al.} protocol \cite{PRL-counterfactuladsqc} 
all other existing protocols of DSQC are based on conjugate coding.
(2) In the proposed protocol actual information encoded quantum state
never propagates through the transmission channel. (3) Whereas the
security of conventional QSDC and DSQC protocols, like that of BB84-class
QKD protocols, is based on conjugate coding, the security
of the present GV-type DSQC protocols is based on monogamy of entanglement.

Present work provides a protocol of DSQC that is similar to the protocol of Salih \emph{et al.} in the sense that its security does not arise from noncommutativity and message qubits are not transmitted. Interestingly,
criticism of Vaidman is not applicable to the present protocol as transmission of non-message qubits happens in the present protocol and thus it works in a weaker condition than that claimed by Salih \emph{et al.} \cite{PRL-counterfactuladsqc}. Further, recently proposed Zhang
et al. \cite{QSDC new-cluster} protocol may be viewed as a  special case of our protocol. This point is explicitly  illustrated through Example
2 of Table \ref{tab:Intersting-quantum-states} where we show that the 4-qubit cluster
state used by Zhang et al. is a special case of the quantum state
(\ref{eq:state of interest}).  The state
described here and the proposed protocol is much more general. In
Table \ref{tab:Intersting-quantum-states}, we have provided 8 examples
of quantum state of the form (\ref{eq:state of interest}) that may
be used for implementation of DSQC using the protocol presented here.
The list can be extended arbitrarily as we can generate infinitely
many quantum states of the form (\ref{eq:state of interest}). However,
a protocol is useful if and only if that can be implemented using
the quantum states that can be generated experimentally using the
contemporary facilities. Interestingly, all the states listed in Table
\ref{tab:Intersting-quantum-states} can be generated in modern laboratories
and their generations are reported in recent past. 

The protocol is shown to be unconditionally secure. It is also noted
that the security of the protocol arises from non-realistic nature
of quantum mechanics. Interestingly it is found that the protocol
is not maximally efficient as far as its qubit efficiency is considered.
This interesting and completely orthogonal state based protocol is
expected to be of much use in all future experimental developments
as it provides a wide choice of quantum states to experimentalists.

\textbf{Acknowledgments:} AP thanks Department of Science and Technology
(DST), India for support provided through the DST project No. SR/S2/LOP-0012/2010
and he also acknowledges the supports received from the projects CZ.1.05/2.1.00/03.0058
and CZ.1.07/2.3.00/20.0017 of the Ministry of Education, Youth and
Sports of the Czech Republic. Authors also thank Dr. R. Srikanth for
some useful technical discussions.

\end{document}